# Does observability amplify sensitivity to moral frames? Evaluating a reputation-based account of moral preferences

**Valerio Capraro[1*], Jillian J. Jordan[2*], and Ben M. Tappin[3*]**

[1]Middlesex University London
[2]Kellogg School of Management, Northwestern University
[3]Sloan School of Management, Massachusetts Institute of Technology

*These authors contributed equally to this work.

Contact: VC, v.capraro@mdx.ac.uk; JJJ, jillian.jordan@kellogg.northwestern.edu; BMT, benmtappin@googlemail.com.

**Abstract**

A growing body of work suggests that people are sensitive to moral framing in economic games involving prosociality, suggesting that people hold moral preferences for doing the "right thing". What gives rise to these preferences? Here, we evaluate the explanatory power of a reputation-based account, which proposes that people respond to moral frames because they are motivated to look good in the eyes of others. Across two pre-registered experiments (total $N = 3,610$), we investigated whether reputational incentives amplify sensitivity to framing effects. Both experiments manipulated (i) whether moral or neutral framing was used to describe a Trade-Off Game (in which participants chose between prioritizing equality or efficiency) and (ii) whether Trade-Off Game choices were observable to a social partner in a subsequent Trust Game. We find that framing effects are relatively *insensitive* to reputational incentives: observability did not significantly amplify sensitivity to moral framing. However, our results are not inconsistent with the possibility that observability has some amplification effect; quantitatively, the observed framing effect was 74% as large when decisions were private as when they were observable. These results suggest that moral frames may tap into moral preferences that are relatively deeply internalized, and that power of moral frames to promote prosociality may not be strongly enhanced by making the morally-framed behavior observable to others.

# Introduction

Humans are exceptionally prosocial. As one important illustration, many canonical experiments using economic games have shown that some people act prosocially even in one-shot anonymous interactions, when no direct or indirect benefits seem to be at play (see Camerer (2011) for a review). Understanding what drives such prosociality is essential for the success of human societies, given that we face global challenges such as resource depletion, climate change, social and economic inequalities, and pandemics (Hardin, 1968; Trivers, 1971; Axelrod & Hamilton, 1981; Nowak, 2006; Perc et al. 2017; Van Bavel et al. 2020).

One classic explanation for human prosociality—and in particular, for prosociality in one-shot anonymous economic games—is that people have outcome-based social preferences (Levine, 1998; Fehr & Schmidt, 1999; Bolton & Ockenfels, 2000; Charness & Rabin, 2002; Engelmann & Strobel, 2002). This explanation poses that people care about the monetary payoffs and outcomes of others, not just themselves (e.g., through preferences for others to receive resources, or for allocations of resources to be equal). Recent research, however, has suggested that beyond having outcome-based social preferences, people also have "moral preferences" for doing the right thing.

**Moral Frames and Moral Preferences**

Over the last decade, a series of papers has revealed that people's decisions in one-shot economic games depend on how the available choices are labelled, especially when the labels activate moral concerns. For example, Krupka and Weber (2013) found that the rate of altruism in the dictator game depends on how the game is framed: dictators tend to be more altruistic in the "take frame" compared to the "give frame", despite the fact that the payoff outcomes for all parties are identical in both versions of the game. And this effect seems to be driven by moral considerations: dictators tend to rate "taking" from the recipient as more socially inappropriate than "not giving to" the recipient.

Furthermore, this type of effect has been documented across a wide range of economic game contexts. Framing effects have been documented for six dictator game frames by Capraro and Vanzo (2019). Eriksson et al. (2017) found that rejection rates in the ultimatum game depend on how the available actions are named, such that participants are more willing to reject an offer when doing so is described as "rejection of an offer" versus "reduction of a payoff". And like Krupka and Weber (2013), Eriksson et al. (2017) also found evidence that this framing effect reflected moral considerations: "reducing the proposer's payoff" was rated as morally worse than "rejecting the proposer's offer". Finally, Capraro and Rand (2018) and Tappin and Capraro (2018) found that minor changes in the framing of trade-off games pitting equity against efficiency can massively impact people's decisions, at least when the payoffs are not too large (Huang, Lei, Xu, Yu, & Shi, 2019). And evidence suggests that, again, these framing effects can be explained by a change in the perception of what is the moral thing to do (Capraro & Rand, 2018). Together, these experiments robustly illustrate that beyond having outcome-based social preferences concerning the payoffs of others, people have preferences for doing the "right thing".

These moral preferences provide a useful tool for intervening to promote prosociality. A growing body of work suggests that "moral suasion" (i.e., using frames that make the morality of an action salient) can function to increase socially desirable behavior. An earlier paper by Brañas-Garza (2007) found that telling dictators that "the other person relies on you" increases giving. Dal Bó and Dal Bó (2014) reported that reading about the Golden Rule encourages cooperation in the iterated prisoner's dilemma. Capraro et al. (2019) showed that asking participants "what do you personally think is the right thing to do?", prior to their participation in a dictator game or a prisoner's dilemma, increased prosociality in these games. Bilancini et al. (2020) found that moral suasion decreased ingroup favoritism on average (although it increased ingroup favoritism for a subset of participants).

The effect of moral suasion has also been observed in decisions that have consequences outside the laboratory. For example, Capraro et al. (2019) demonstrated that their moral nudge increases online crowdsourcing of charitable donations to humanitarian organizations by 44%, and Bott, Cappelen, Sørensen and Tungodden (in press) recently observed that letters with moral reminders decrease tax evasion among Norwegian tax-payers.

Thus, previous work has both demonstrated that people readily respond to moral frames—reflecting moral preferences for doing the right thing—and illustrated the potential for interventions to capitalize on moral preferences in order to encourage socially desirable behavior. However, much less is known about *why* people exhibit moral preferences and are sensitive to framing. What underlies the preference to do the "right" thing, and the resulting sensitivity to how choices are framed? Here, we consider this question, both from the perspective of the "ultimate" mechanisms through which moral preferences and framing effects may be adaptive, and the "proximate" mechanisms through which moral preferences and framing effects may be psychologically instantiated.

**A Reputation-Based Explanation for Framing Effects**

One potential explanation for moral preferences and framing effects comes from considering the role of *reputation*. People are strongly motivated to be seen positively by others, and thus reputation is an important driver of prosocial behavior (Barclay, 2004; Boyd & Richerson, 1989; Emler, 1990; Milinski et al. 2002; Nowak & Sigmund, 2005). Supporting this proposal, evidence suggests that removing confidentiality increases public goods contributions (Andreoni & Petrie, 2004; Rege & Telle, 2004) and charity donations (Bereczkei, Birkas, & Kerekes, 2007; Bracha & Meier, 2009; Lacetera & Macis, 2010). (See Bradley, Lawrence and Ferguson (2018) for a recent meta-analysis.) Thus, one plausible reason for why people are sensitive to moral frames is that frames provide information about what will be seen as moral by *other* people; and, thus, responding to frames can confer reputational benefits. In other words, framing effects may draw on the reputation-based drive to appear moral in the eyes of others.

Yet, we know that framing effects can occur even in contexts where nobody is watching (e.g., one-shot anonymous economic game experiments). On its face, this observation is inconsistent with the proposal that framing effects draw on reputation motives: when nobody is watching, people are unlikely to be explicitly concerned with looking good in the eyes of others. However, evidence suggests that some people rely on the heuristic that reputation is typically at stake and use this heuristic even in one-shot game experiments where nobody is watching (Jordan & Rand, 2019). Such a heuristic might function to avoid the cognitive costs of computing whether an interaction is likely to be observed (Delton, Krasnow, Cosmides, & Tooby, 2011; Kahneman, 2011; Rand, Tomlin, Bear, Ludvig, & Cohen, 2017) or the social costs of performing such a computation—given that carefully calculating whether to cooperate or not can be perceived negatively by observers (Capraro & Kuilder, 2016; Critcher, Inbar, &

Pizarro, 2013; Hoffman, Yoeli, & Nowak, 2015; Jordan, Hoffman, Nowak, & Rand, 2016). Furthermore, people might default towards behaving as if they are being watched as an "error management" strategy, designed to avoid the costs of mistaking a situation in which reputation is at stake for an anonymous situation (Delton, Krasnow, Cosmides, & Tooby, 2011).

Accordingly, it is plausible that framing effects *do* draw on reputation motives, but nonetheless occur in anonymous interactions because people are sensitive to reputation cues even when nobody is watching. Of course, a reputation-based explanation for framing effects might still predict that sensitivity to framing increases meaningfully in contexts where reputation *is* actually at stake, and people thus have explicit reputation motives. For example, Jordan and Rand (2019) found that the presence of reputational incentives can increase sensitivity to reputation cues among deliberative individuals (who are less likely to heuristically assume that reputation is at stake even in anonymous interactions). Thus, to the extent that framing effects draw on reputation concerns, we might expect them to increase when explicit reputational incentives are introduced.

Here, we evaluate this prediction by investigating the extent to which sensitivity to framing is itself sensitive to whether or not reputation is at stake. This inquiry has important practical implications: if sensitivity to framing is amplified substantially by the presence of explicit reputational incentives, it suggests that there is substantial value in targeting framing-based interventions at contexts where reputation is at stake. Furthermore, investigating the effect of reputational incentives on sensitivity to framing can shed light on the mechanisms underlying framing effects. Specifically, such an investigation can shed light on the extent to which framing effects draw on psychological processes that are context-sensitive (and thus increase substantially when explicit reputational incentives are introduced) versus deeply internalized (and thus relatively insensitive to whether we are being watched).

**Our contribution**

In this paper, we investigate the influence of reputational incentives on sensitivity to moral frames in the context of the Trade-Off Game (TOG), an economic decision problem that forces decision-makers to decide between competing values of equality and efficiency. We chose to investigate decisions in the TOG because there is a robust body of evidence that, in experiments where reputation is *not* at stake, moral framing has a substantial influence on TOG decisions (Capraro & Rand, 2018; Tappin & Capraro, 2018; Capraro, Rodriguez-Lara, & Ruiz-Martos, 2020; Huang, Lei, Xu, Yu, & Shi, 2019).

We thus conducted two pre-registered experiments in which participants (i) decided how to behave in the TOG, and then (ii) participated in an economic Trust Game with another player. As described further below, participants participated in the Trust Game in the role of the "Trustee", and thus faced incentives to appear trustworthy in the eyes of the other player.

Our experiments employed a two-by-two, between-subjects design in which we manipulated (i) the framing used to describe participants' TOG decisions (neutral frame vs. moral frame) and (ii) whether or not participants' TOG decisions were observable to the other player in the Trust Game (private condition vs. public condition). This design allows us to investigate the extent to which activating reputation concerns (by making TOG decisions observable) increases sensitivity to moral framing.

In the Trust Game, we also measured the extent to which focal participants actually behaved in a trustworthy way towards the other player. Thus, our design also allows us to ask an additional question: to what extent do individuals who make morally framed choices in the TOG behave in a more trustworthy way in the Trust Game? Previous work has found that individuals who make morally framed choices in the TOG tend to behave more prosocially in subsequent one-shot anonymous economic games like the Dictator Game and Prisoner's Dilemma (Capraro & Rand, 2018; Tappin & Capraro, 2018). These results further support the proposal that sensitivity to framing reflects the prosocial preference to do the right thing. And they lend credulity to the reputation-based hypothesis we explore in our primary analyses: Insofar as morally-framed choices are a meaningful signal of prosociality, it makes sense that the desire to appear prosocial in the eyes of others might serve to amplify sensitivity to moral frames. Thus, as a secondary question of interest in our paper, we sought to replicate the association between morally-framed choices and prosociality in the context of our paradigm. To this end, we investigated whether selecting the morally-framed choice in the TOG would serve to signal trustworthiness in the Trust Game.

## Method

The study designs, hypotheses, and analysis plans were pre-registered. The protocols are online at https://aspredicted.org/yz4gk.pdf (Study 1) and https://aspredicted.org/zz5my.pdf (Study 2). Owing to their close similarity, we present the results of these two studies together.

**Samples**

Participants were recruited via Amazon's Mechanical Turk (AMT, Paolacci, Chandler & Ipeirotis, 2010; Horton, Rand & Zeckhauser, 2011; Paolacci & Chandler, 2014; Arechar, Gächter & Molleman, 2018). The sample sizes and payment information are reported below (subheading "Study 1 and Study 2").

**Procedure and Design**

Participants were randomly assigned to one of four conditions, corresponding to a two-by-two, between-subjects design in which we manipulated Choice Frame (Neutral, Moral) and Choice Observability (Public, Private). We describe each of these manipulations in the context of our experimental procedure below.

**Trade-Off Game.** Participants began the study by learning that they would be playing a game called the "Trade-Off Game" (TOG). The TOG instructions described to participants that they would be playing as "Player A", and that they would be matched with two other players, "Player B" and "Player C". As Player As, participants were asked to choose between one of two options: Option 1 (referred to by us in this paper as the equitable option) and Option 2 (referred to by us in this paper as the efficient option). Option 1 involved providing 13 cents to each of the three players (Player's A, B, and C). In contrast, Option 2 involved providing 13 cents to Player A, 23 cents to Player B and 13 cents to Player C. Thus, Option 2 provided a larger total payoff for the group (and was thus desirable from an efficiency standpoint), but also conferred an unequal distribution of money to group members (and was thus undesirable from an equality standpoint).

In the Neutral Frame condition, we labelled the options simply as "Option 1" and "Option 2". In contrast, in the Moral Frame condition, we labelled Option 1 (i.e., the option that values equality) as the "fair" option. Previous research has established that, in this Moral Frame condition, the vast majority of participants do think that selecting Option 1 is the morally right thing to do (Capraro & Rand, 2018).

After reading the TOG instructions, all participants were asked two comprehension questions to assess their understanding of the task.

**Trust Game.** After completing the TOG comprehension questions, participants were informed that they would be playing another game: The Trust Game (TG). All participants were informed they would be playing the TG in the role of the "Receiver" (i.e., the Trustee), and that a new player (who was *not* involved in the TOG) would be playing the TG in the role of the "Sender" (i.e., the Trustor). Participants were then told that the Sender would start with 30 cents and choose how many cents, if any, to send to them. Each cent sent would be tripled before being given to the participant, who would then get to decide how much, if anything, to return to the Sender.

In the *Public* condition, participants were told that the Sender would learn about the existence of the TOG and find out about their TOG decision. In contrast, in the *Private* condition, participants were told that the Sender would *not* learn about the existence of the TOG or find out about their TOG decision. Thus, in the Public condition, participants knew that their Trust Game partner would observe their TOG decision before deciding how much money to entrust them with, whereas participants in the Private condition knew that their TOG decision would not be observed.

Following the TG instructions, participants were asked four comprehension questions to assess their understanding of the task.

**Game Decisions.** After completing the TG comprehension questions, all participants made their game decisions. First, they decided whether to make the equitable or efficient choice in the TOG, and second, they decided how much to return (as a proportion of whatever amount they would be sent) to their partner in the TG. In the TOG decision screen, participants were reminded whether their choice would (Public condition) or would not (Private condition) be revealed to the Sender. The TG decision screen was the same for all the participants. Verbatim instructions for both the TOG and TG are reported in the Supplementary Information. Finally, after making their decisions, participants completed a demographic questionnaire.

**Study 1 and Study 2.** As described, we collected data in two pre-registered studies. In Study 1, the participation payment was 50 cents and we pre-registered a target sample size of N=800 participants. In Study 2, we pre-registered a much larger target sample size of N=3,000. The procedure for Study 2 was identical to the procedure for Study 1, with the exception that (i) in Study 2 the participation payment was 30 cents (instead of 50 cents) and we correspondingly shortened the demographic questionnaire to reduce the length of the survey, and (ii) in Study 2 the stakes of both economic games were lower. Specifically, the stakes in the TOG were [5,5,5] vs. [5,10,5] cents (instead of [13,13,13] vs. [13,23,13]), and, in the TG, the Sender started with 10 cents (instead of 30 cents).

**Disclosure statement:** In these studies, we report all measures, manipulations and exclusions.

**Sensitivity power analysis:** We conducted sensitivity power analyses for our primary test (Hypothesis 1) via simulations. The script to reproduce our simulations is available online on

https://osf.io/45cey/. For all simulations, we assume a moral framing effect in the private condition equal to an increase of 15% percentage points and a baseline of 50% (i.e., neutral-private condition proportion = 0.50; moral-private condition proportion = 0.65). Accordingly, Study 1 (N = 800) had 80% power to detect an interaction between the moral frame and public condition assuming that the effect of the moral frame in public was approximately double the size of that observed in private: neutral-public condition proportion = 0.50; moral-public condition proportion = 0.82. Owing to its larger sample size (N = 2810), Study 2 had 80% power to detect a smaller increase in the moral frame effect in public: neutral-public condition proportion = 0.50, moral-public condition proportion = 0.75. The pooled data from Studies 1 and 2 had approximately 80% power to detect a slightly smaller effect: neutral-public condition proportion = 0.50, moral-public condition proportion = 0.74.

**Ethics statement.** For all the sessions, relevant ethical guidelines were followed. This research was approved by the MIT IRB, COUHES Protocol #: 18066392996A004.

## Results

Data and analysis code to reproduce all results is available online at https://osf.io/45cey/. For all analyses, we report results among all participants (preregistered primary analyses) and among participants who correctly answered all comprehension questions about both economic games (preregistered exploratory analyses: briefly described at the end of this section, but reported in full in the SI).

We preregistered two data exclusion criteria: We exclude responses from participants who did not complete all of the primary measures prior to the demographics (because these participants by definition could not be included in the preregistered analyses). We also exclude those who had duplicate IP addresses or AMT worker IDs, retaining the first response only (determined by the start date). Our final sample size is thus N=3610 (Study 1 N=800, Study 2 N=2810). Our preregistered analyses address two key questions.

**Analysis 1: Does observability increase sensitivity to moral framing in the TOG?**

Our first and primary analysis asks whether people are more sensitive to the moral framing manipulation when their choice is observable (versus unobservable) by their partner in the subsequent Trust Game. Thus, we test the two-way interaction between framing (neutral, moral) and observability (private, public) on TOG choices. The relevant raw data are displayed in Figure 1A, partitioned by study.

We analyze these data using binomial logistic regression. The results for each study are reported in Table 1 (Analysis 1). Replicating past work and illustrating participants' preferences to behave morally, we find consistent evidence across studies for a main effect of the moral frame: Participants are more likely to choose the equitable (vs. efficient) option in the TOG when that choice is framed as the "fair" option than when there is no such frame. This effect can be seen in Figure 1A by comparing results in the moral versus neutral frame conditions. In Table 2 (Analysis 1), we pool the data across the studies—including a dummy variable to indicate study-specific differences in the intercept—to obtain a more precise estimate of the moral framing estimates.

Next, we turn to our key question: Does observability increase sensitivity to moral framing? We do not find strong evidence that it does. That is, our results imply that participants are sensitive to moral framing both in *private* (where reputational incentives are absent, and participants are unlikely to be explicitly motivated by reputation concerns) and in *public* (where reputational incentives are present, and participants may be explicitly motivated by reputation concerns). And while the observed framing effect is directionally larger in public than in private (as illustrated in Figure 1A), we do not find evidence for a significant amplification in the presence of reputational incentives. Table 1 (Analysis 1) reveals that we do not find a significant interaction between our moral frame and observability manipulations in either of our studies individually, and Table 2 (Analysis 2) reveals that pooling data across studies to maximize precision gives the same result.

Thus, our studies fail to provide strong evidence that observability increases sensitivity to moral framing, and thus suggest that moral framing may be similarly effective in the absence versus presence of reputational incentives. To provide a quantitative sense of this conclusion, we note that, pooling data across both studies, we find that the moral framing effect in the Private condition is approximately 74% as large as the moral framing effect in the Public condition.

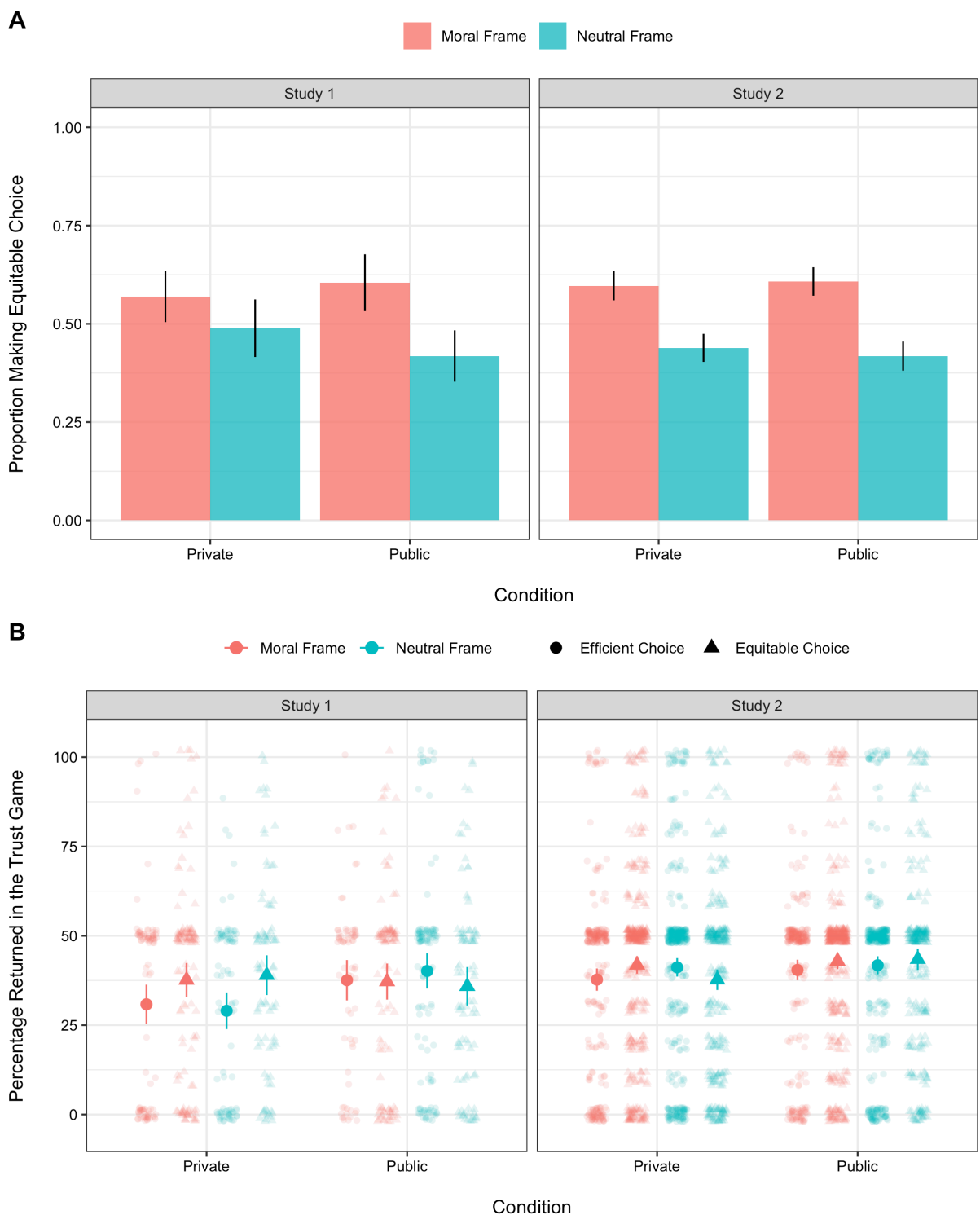

**Figure 1. Data from Studies 1 and 2.** (A) Proportion of participants choosing to split the money equitably in the Trade-Off Game as a function of observability (Private, Public) and choice frame (Moral, Neutral)**.** (B) Percentage of money returned in the Trust Game as a function of observability (Private, Public), choice frame (Moral, Neutral), and choice made in the Trade-Off Game (Efficient, Equitable). Large solid points are the means in each cell. Smaller faded points are the raw data with slight jitter to aid visibility. (A, B) Error bars are 95% CI. N = 800 in Study 1; N = 2810 in Study 2.

**Table 1.** Results from Analyses 1 and 2 in each of Studies 1 and 2.

| Analysis | Study | Term | Estimate | 95 LL | 95 UL | p-value |
|---|---|---|---|---|---|---|
| 1 | | | | | | |
| | 1 | (Intercept) | 0.96 | 0.71 | 1.28 | .766 |
| | | Moral Frame | 1.38 | 0.93 | 2.05 | .107 |
| | | Public Condition | 0.75 | 0.51 | 1.12 | .158 |
| | | Moral x Public | 1.54 | 0.87 | 2.70 | .135 |
| | 2 | (Intercept) | 0.78 | 0.68 | 0.90 | .001 |
| | | Moral Frame | 1.89 | 1.53 | 2.34 | < .001 |
| | | Public Condition | 0.92 | 0.74 | 1.13 | .423 |
| | | Moral x Public | 1.14 | 0.84 | 1.54 | .391 |
| 2a | | | | | | |
| | 1 | (Intercept) | 30.83 | 25.36 | 36.31 | < .001 |
| | | Equitable Choice | 6.80 | -0.45 | 14.06 | .067 |
| | 2 | (Intercept) | 37.75 | 34.71 | 40.78 | < .001 |
| | | Equitable Choice | 4.10 | 0.17 | 8.03 | .041 |
| 2b | | | | | | |
| | 1 | (Intercept) | 29.02 | 23.57 | 34.48 | < .001 |
| | | Equitable Choice | 9.96 | 2.15 | 17.76 | .013 |
| | | Moral Frame | 1.81 | -5.82 | 9.45 | .642 |
| | | Equitable x Moral | -3.15 | -13.69 | 7.38 | .558 |
| | 2 | (Intercept) | 41.17 | 38.65 | 43.69 | < .001 |
| | | Equitable Choice | -3.47 | -7.27 | 0.34 | .074 |
| | | Moral Frame | -3.43 | -7.43 | 0.57 | .093 |
| | | Equitable x Moral | 7.56 | 2.03 | 13.1 | .007 |

*Note*. Analysis 1 model estimates and CIs are odds ratios. Analysis 2a and 2b are OLS estimates. LL = Lower Limit. UL = Upper Limit.

**Table 2.** Results from Analyses 1 and 2 pooling the data across Studies 1 and 2.

| Analysis | Term | Estimate | 95 LL | 95 UL | p-value |
| --- | --- | --- | --- | --- | --- |
| 1 | | | | | |
| | (Intercept) | 0.82 | 0.69 | 0.99 | .037 |
| | Moral Frame | 1.77 | 1.47 | 2.13 | < .001 |
| | Public Condition | 0.88 | 0.73 | 1.06 | .184 |
| | Study 2 Dummy | 0.99 | 0.84 | 1.16 | .854 |
| | Moral x Public | 1.22 | 0.93 | 1.59 | .145 |
| 2a | | | | | |
| | (Intercept) | 31.99 | 28.04 | 35.94 | < .001 |
| | Equitable Choice | 4.77 | 1.31 | 8.23 | .007 |
| | Study 2 Dummy | 5.35 | 1.40 | 9.30 | .008 |
| 2b | | | | | |
| | (Intercept) | 34.45 | 31.13 | 37.77 | < .001 |
| | Equitable Choice | -0.83 | -4.25 | 2.59 | .635 |
| | Moral Frame | -2.59 | -6.13 | 0.95 | .152 |
| | Study 2 Dummy | 5.52 | 2.59 | 8.45 | < .001 |
| | Equitable x Moral | 5.60 | 0.70 | 10.50 | .025 |

*Note*. Analysis 1 model estimates and CIs are odds ratios. Analysis 2a and 2b are OLS estimates. LL = Lower Limit. UL = Upper Limit.

**Analysis 2: Do Morally-Framed Decisions Signal Trustworthiness?**

Our second analysis asks whether morally-framed decisions signal trustworthiness, and is broken into two parts. To address this question, we first ask in Analysis 2a whether participants who choose the morally-framed option (i.e., the equitable choice) in the TOG behave in a more trustworthy manner in the subsequent Trust Game. Then, in Analysis 2b, we turn to asking whether any such association depends on the presence of the moral frame, or is instead explained by a general association between equitable choices and trustworthiness (that holds irrespective of framing). The relevant raw data are shown in Figure 1B.

In these analyses, we investigate the association between morally-framed choices and trustworthiness in the *Private* condition of our paradigm, in which reputational incentives were absent and thus sensitivity to moral framing can only reflect "pure" preferences for doing the right thing. This approach has the advantage of allowing us to straightforwardly compare our results to previous work investigating the association between morally-framed choices and prosociality, which has employed one-shot anonymous games that are analogous to our Private condition (Capraro & Rand, 2018; Tappin & Capraro, 2018). These analyses were thus preregistered as primary analyses because we had clear theoretical predictions. Because we did not have clear theoretical predictions about how the Public condition might affect the association between morally-framed choices and trustworthiness, we preregistered analyses on the Public condition as exploratory, and we prioritize brevity by not additionally reporting those analyses here. However, we note that interested readers can get a visual sense of the relevant associations from Figure 1B, and we report formal results from these exploratory analyses on the Public condition specifically in the SI.

**Analysis 2a.** We first examine the trustworthiness of participants in the Private condition who received the moral frame in the TOG. This functions as a basic test of whether participants who choose the equitable option behaved in a more trustworthy manner than participants who choose the efficient option, when the equitable option is framed as morally good and decisions are private. Understood with respect to Figure 1B, this test compares the first two cells of the Private condition. We analyze these data using OLS regression. The results for each study are reported in Table 1 (Analysis 2a), and the pooled estimate across both studies is reported in Table 2 (Analysis 2a).

We find some evidence that participants who chose the equitable option returned more in the subsequent Trust Game than participants who chose the efficient option: the estimates in both studies are positive, and the p-values on the estimates are .067 (Study 1) and .041 (Study 2). Thus, we find some evidence that equitable choices predict trustworthiness—when such choices are framed as morally good and all decisions are private.

**Analysis 2b.** We next proceed to ask: Is the association identified in Analysis 2a observed only when the equitable choice is framed as morally good, or does it instead reflect a general association between equitable choices and trustworthiness (that holds irrespective of framing)?

To address this question, we investigate the two-way interaction between framing (neutral, moral) and TOG choice (equitable, efficient) on trustworthiness in the TG. Understood with respect to Figure 1B, this test compares the difference between the first two cells of the Private condition with the difference between the latter two cells of the Private condition. We analyze these data using OLS regression. The results for each study are reported in Table 1 (Analysis 2b), and the pooled estimate across both studies is reported in Table 2 (Analysis 2b).

We find some evidence that participants who chose the equitable (versus efficient) option returned more in the subsequent Trust Game *only* when the equitable option was framed as morally good. Specifically, while there is no support for this pattern (even directionally) in Study 1, the evidence does support it in Study 2, and in the pooled estimate across both studies (as indicated by the significant positive interaction). Thus, we find some evidence that, when decisions are private, equitable choices specifically predict trustworthiness when they are framed as morally good.

Together, the results of Analysis 2 are thus broadly supportive of the proposal that, in a context where reputational incentives are absent—and thus sensitivity to moral framing can only reflect "pure" moral preferences—morally-framed choices possess some value as a signal of trustworthiness.

**Exploratory Analyses Among Perfect Comprehenders**

Finally, as mentioned, we also conducted the above analyses (i.e., 1, 2a, and 2b) on a restricted sample of participants: those who correctly answered all the comprehension questions for the TOG and TG, dubbed "perfect comprehenders". The results are reported in figures and tables in the SI. The main result of note is that, in Study 2, Analysis 1 on the perfect comprehenders, we *do* observe a statistically significant interaction between moral frame and public condition (p = .029), reflecting that the moral framing effect on TOG choice is larger in public, and this difference is statistically distinguishable at the .05 level.

## Discussion

A growing body of research has illustrated that people are sensitive to framing when making moral decisions. This research suggests that beyond having outcome-based social preferences (e.g., for others to receive resources, or for resource allocations to be equal), people also have preferences for doing the "right" thing. In this work, we have investigated the extent to which reputational incentives serve to amplify sensitivity to framing effects, and thus the extent to which the moral preferences underlying framing effects are sensitive to whether reputation is at stake. And we have provided evidence that framing effects are relatively *insensitive* to the presence of reputational incentives. In our studies, participants responded to frames even when their decisions were completely private, and we did not find reliable evidence that introducing reputational incentives increases sensitivity to framing.

Importantly, our results are not inconsistent with the possibility that reputational incentives do have some power to increase sensitivity to framing. Directionally, we did find larger framing effects when reputation was at stake; recall that across our two studies, the framing effect was 74% as large in the Private condition as in the Public condition. Furthermore, as briefly noted in our results, in our exploratory analyses of participants who correctly answered all comprehension questions, we do find some evidence that reputational incentives amplify sensitivity to framing. Yet in our pre-registered primary analyses of all participants, across two studies with a total $N$ of 3,610 (providing approximately 80% power to detect a minimum increase in sensitivity to framing of nine percentage points), we were not able to detect significant amplification. We thus see our results as suggesting that framing effects are *relatively* insensitive to the presence of reputational incentives, and that reputational incentives do not *strongly* amplify sensitivity to moral frames.

Our studies do, however, lend more support to the proposal that morally-framed choices serve as a reliable signal of trustworthiness. In our secondary analyses, we found evidence that equitable decisions in the Tradeoff Game predict returning in the Trust Game—but only when such choices are framed as morally good. These analyses are consistent with previous evidence that morally-framed decisions predict prosocial behavior (Capraro & Rand, 2018; Tappin & Capraro, 2018), and serve to further bolster the proposal that sensitivity to moral framing reflects the prosocial preference to do the right thing. They also lend credulity to the reputation-based hypothesis we investigated in our primary analyses by suggesting that sensitivity to moral framing may be an effective reputational strategy (insofar as observers are aware that morally-framed choices signal prosociality). Thus, it is all the more notable that sensitivity to framing is relatively insensitive to the presence of reputational incentives.

Our results have important implications for interventions that draw on framing effects to encourage socially desirable behavior. They suggest that such interventions can be successful even when behavior is not observable to others and thus reputation is not at stake—and in fact, that the efficacy of framing effects is not strongly enhanced by making behavior observable. Thus, our results suggest that targeting contexts where reputation is at stake is not an especially important priority for individuals seeking to maximize the impact of framing-based interventions. This conclusion provides an optimistic view of the potential of framing-based interventions, given that there may be many contexts in which it is difficult to make behavior observable but yet possible to frame a decision in a way that encourages prosociality.

Our results also have important implications for the psychology underlying framing effects. Specifically, they suggest that framing effects draw on preferences that are relatively deeply internalized (and thus are relatively insensitive to the presence of reputational incentives). This observation could reflect that reputation systems do *not* explain why people are sensitive to framing. In other words, our results could reflect that people react to framing not because they see frames as conveying information about what others will see as moral, but rather because they see frames as conveying information about what actually is moral—and frames tap into a general and deeply internalized desire to behave morally (and/or think of the self as behaving morally) (Ariely, Bracha, & Meier, 2009; Mazar, Amir, & Ariely, 2008). Such a desire would serve to encourage cooperative and moral behavior, and thus could be supported by many ultimate mechanisms beyond reputation systems that can give rise to cooperative and moral behavior (e.g., direct reciprocity, kin selection, institutions, and cultural group selection) (Boyd & Richerson, 2009; Henrich, 2006; Nowak, 2006; Trivers, 1971).

Alternatively, however, our results could be compatible with the proposal that reputation systems explain why people are sensitive to framing. Specifically, our results could reflect that people do react to framing because frames convey information about what others will see as moral—and consequently do tap into a more specific desire to appear moral in the eyes of others that is ultimately supported specifically by reputation systems. However, insofar as reputation is the primary driver of framing effects, our results suggest that the relevant reputation motives are strongly activated even in anonymous experiments, such that introducing reputational incentives does not substantially amplify framing effects. This possibility is consistent with the proposal that even when nobody is watching and people are not *explicitly* concerned with their reputations, they may nonetheless be *implicitly* motivated to appear virtuous—reflecting the use of heuristics (specifically, the heuristic that reputation is typically at stake) (Jordan & Rand, 2019) and/or error management strategies (specifically, strategies designed to avoid the costs of mistaking a situation in which reputation is at stake for an anonymous situation) (Delton et al., 2011).

Regardless, however, our results suggest that framing effects function by drawing on preferences that are not highly sensitive to the presence of reputational incentives. And consequently, they suggest that a sensitivity to framing effects—and the underlying preference to behave morally—may be thought of as a core and deeply internalized feature of our moral psychology. This conclusion implies that we might expect framing effects to be relatively robust and invariant across contexts, a prediction that should be probed further in future research. In particular, an interesting question is whether introducing reputational incentives may be a more effective strategy for amplifying the power of framing effects in contexts where the baseline sensitivity to framing is relatively low (e.g., because making the morally framed decision is particularly costly or difficult).